# Novel BiS$_2$-based layered superconductor Bi$_4$O$_4$S$_3$

Yoshikazu Mizuguchi[1,4,*], Hiroshi Fujihisa[2], Yoshito Gotoh[2], Katsuhiro Suzuki[3], Hidetomo Usui[3], Kazuhiko Kuroki[3], Satoshi Demura[4], Yoshihiko Takano[4], Hiroki Izawa[1], Osuke Miura[1]

1. Department of Electrical and Electronic Engineering, Tokyo Metropolitan University, 1-1, Minami-osawa, Hachioji, Tokyo 192-0397, Japan.
2. National Institute of Advanced Industrial Science and Technology (AIST), Tsukuba Central 5, 1-1-1, Higashi, Tsukuba 305-8565, Japan.
3. Department of Engineering Science, The University of Electro-Communications, 1-5-1 Chofugaoka, Chofu, Tokyo 182-8585, Japan.
4. National Institute for Materials Science, 1-2-1, Sengen, Tsukuba 305-0047, Japan.

[Contact information]
*Corresponding author: Yoshikazu Mizuguchi
  Affiliation: Tokyo Metropolitan University
  E-mail: mizugu@tmu.ac.jp
  Address: 1-1, Minami-osawa, Hachioji, 192-0397, Japan

Exotic superconductivity has often been discovered in materials with a layered (two-dimensional) crystal structure. The low dimensionality can affect the electronic structure and can realize high transition temperatures ($T_c$) and/or unconventional superconductivity mechanisms. As standard examples, we now have two types of high-$T_c$ superconductors. The first group is the Cu-oxide superconductors whose crystal structure is basically composed of a stacking of spacer (blocking) layers and superconducting $CuO_2$ layers.[1-4] The second group is the Fe-based superconductors which also possess a stacking structure of spacer layers and superconducting $Fe_2An_2$ (An = P, As, Se, Te) layers.[5-13] In both systems, dramatic enhancements of $T_c$ are achieved by optimizing the spacer layer structure, for instance, a variety of composing elements, spacer thickness, and carrier doping levels with respect to the superconducting layers. In this respect, to realize higher-$T_c$ superconductivity, other than Cu-oxide and Fe-based superconductors, the discovery of a new prototype of layered superconductors needs to be achieved. Here we show superconductivity in a new bismuth-oxysulfide layered compound $Bi_4O_4S_3$. Crystal structure analysis indicates that this superconductor has a layered structure composed of stacking of $Bi_4O_4(SO_4)_{1-x}$ and $Bi_2S_4$ layers; the parent compound ($x = 0$) is $Bi_6O_8S_5$. Band calculation suggests that $Bi_4O_4S_3$ ($x = 0.5$) is metallic while $Bi_6O_8S_5$ ($x = 0$) is a band insulator with $Bi^{3+}$. Furthermore, the Fermi level for $Bi_4O_4S_3$ is just on the peak position of the partial density of states of the Bi 6p orbital within the $BiS_2$ layer. The $BiS_2$ layer is a basic structure which provides another universality class for layered superconducting family, and this opens up a new field in the physics and chemistry of low-dimensional superconductors.

In the 100 years since the discovery of superconductivity, many superconductors have been discovered. In the early stages, new superconductors had been found in metallic compounds such as Nb-Ti alloy and Nb$_3$Sn. In 1986, Bednorz and Müller reported high-$T_c$ superconductivity in (La,Ba)$_2$CuO$_4$ at 30K.[1] After this discovery, a dramatic increase of $T_c$ was achieved by optimizing the spacer layers and the number of CuO$_2$ planes, for example in YBa$_2$Cu$_3$O$_x$ ($T_c$ > 90 K), Bi$_2$Sr$_2$Ca$_2$Cu$_3$O$_x$ ($T_c$ > 110 K) and HgBa$_2$Ca$_2$Cu$_3$O$_x$ ($T_c$ > 130 K).[2-4] The high-$T_c$ Cu-oxide (cuprate) superconductors provided us with some strategies to explore new high-$T_c$ superconductors, such as layered (low-dimensional) crystal structures, insulator-metal boundaries, electronic correlations and the correlation between magnetism and superconductivity.

After the discovery of the cuprate family, some layered superconductors were discovered. Layered magnesium diboride MgB$_2$ exhibits a comparably high $T_c$ of 39K although it is a simple binary compound.[14] Layered nitride HfNCl and ZrNCl are band insulators. However, when electrons are doped by intercalation of ions into the interlayer site, an insulator-metal transition occurs and a superconductivity of $T_c$ > 25 K is achieved.[15,16] As in the Hf-nitride superconductor, superconductivity in layered Co-oxide Na$_x$CoO$_2$ is achieved by intercalation of H$_3$O$^+$ ions into the interlayer site.[17] Most recently, high-$T_c$ superconductivity was discovered in Fe-based compounds. Kamihara and Hosono et al. reported that the antiferromagnetic metal LaFeAsO became superconducting ($T_c$ > 26 K) when the O was partially substituted by F.[5] Soon after the discovery, SmFeAsO$_{1-x}$F$_x$ was reported to be a superconductor with $T_c$ > 55 K.[6,7] To date, over 50 Fe-based superconductors with a layered structure, basically composed of superconducting layers of Fe-pnictogen (Fe$_2$As$_2$ or Fe$_2$P$_2$) or Fe-chalcogen (Fe$_2$Se$_2$ or Fe$_2$Te$_2$) layers and spacer layers, have been discovered.[8-13] As introduced here, discovery of a new layered superconductor provides us with a new research field on physics and chemistry. If a new superconducting layer was discovered, we have a challenge for increasing $T_c$ or exploring new superconductors by changing the spacer layer or alignment of superconducting and spacer layers. In this letter, we show superconductivity in a new layered superconductor bismuth-oxysulfide Bi$_4$O$_4$S$_3$.

Polycrystalline samples of Bi$_4$O$_4$S$_3$ were prepared using a conventional solid state reaction method. Bi$_2$S$_3$, Bi$_2$O$_3$ powders and S grains were ground, pelletized, sealed into an evacuated quartz tube and heated at 510 ºC for 10 h. The product was well-ground, pelletized and annealed at 510 ºC for 10 h in an evacuated quartz tube again.

Firstly, we show the physical properties of the new superconductor Bi$_4$O$_4$S$_3$.

Figure 1a shows the temperature dependence of magnetic susceptibility ($\chi$) for $Bi_4O_4S_3$ from 10 to 2 K. Both zero-field-cooled (ZFC) and field-cooled (FC) data begins to decrease below 6 K as indicated by an arrow in the inset of fig. 1a. A large diamagnetic signal is observed below 4.5K in the ZFC curve. The value of $4\pi\chi$(ZFC) at 2 K exceeds -1, indicating that the shielding volume fraction at 2 K is almost 100 %. Namely, $Bi_4O_4S_3$ is a bulk superconductor. Figure 1b shows the temperature dependence of resistivity from 300 to 2 K for $Bi_4O_4S_3$. Resistivity linearly decreases with cooling and begins to drop below ~8.6 K. An enlarged graph is displayed in the inset of fig. 1b. As indicated by an arrow in the inset, below 8.6 K, a decrease of resistivity, corresponding to the onset of superconducting transition, is clearly observed and resistivity reaches zero below 4.5 K. Temperature dependence of resistivity for $Bi_4O_4S_3$ under magnetic fields up to 5 T is displayed in fig. 1c. The superconducting states are destroyed by applying high magnetic fields. The increase of resistivity with applying magnetic fields would be due to the magnet resistance of impurity phase of Bi. To investigate the temperature - magnetic field phase diagram for superconductivity of $Bi_4O_4S_3$, we plotted $T_c^{onset}$ and $T_c^{zero}$ with the respective applied magnetic fields in fig. 1d. The $T_c^{onset}$ at high magnetic fields was estimated as indicated by an arrow in the inset of fig. 1d. The $T_c^{zero}$ linearly decreases with increasing magnetic field. By a linear extrapolation, the irreversible field $\mu_0H_{irr}(0)$ is estimated to be ~1 T. The upper critical field $\mu_0H_{c2}(0)$ was estimated to be ~21 T using the WHH theory,[18] which gives $\mu_0H_{c2}(0) = -0.69T_c(d\mu_0H_{c2}/dT)|_{Tc}$, with a sloop of $T_c^{onset}$ for $H \geq 1.5$ T. To summarize the physical properties of $Bi_4O_4S_3$, the $T_c^{onset}$ and $T_c^{zero}$ are 8.6 K and 4.5 K at 0 T, respectively. The broadness of the transition is probably caused by an inhomogeneity of the local structure and/or career doping level.

Next, we show the crystal structure obtained by Rietveld refinement of the powder x-ray diffraction (XRD) pattern and density functional theory (DFT) calculations. At first, by indexing the observed peaks except for small peaks of impurity phases of $Bi_2S_3$ and Bi, we obtained the *I*-centered tetragonal lattice with lattice parameters of *a* = 3.9592(1) and *c* = 41.241(1) Å. The extinction rule suggested that the *I*4/*mmm* and *I*-42*m* space groups were the candidates. By fitting the XRD pattern with only Bi atoms at the Bi1, Bi2 and Bi3 sites in Table 1, the reliability factor of $R_{wp}$ = 26.5 % was obtained. Furthermore, by adding S atoms at the S1, S2 and S3 sites in Table 1 to the refinement, $R_{wp}$ = 16.0 % was achieved. Since both models for these space groups are completely identical, we selected higher-symmetric *I*4/*mmm*. Because it is difficult to determine the O sites using only XRD data, we performed structural optimization using DFT calculations. Additionally, we have confirmed the structural stability using molecular dynamics simulation. The atomic positions of Bi and S, determined by the XRD pattern, were only stable when the SO$_4$ ions occupied $z = 0$ and 0.5. By using the structural model, we performed Rietveld refinement. Figure 2a shows the XRD pattern of this sample and the result of Rietveld refinement ($R_{wp}$ = 14.41 %). The obtained atomic parameters and the site occupancy are summarized in Table 1. As shown in Fig. 2b, the XRD pattern is well-fitted at high angle region. A schematic image of the final crystal structure is shown in Fig. 2c. The layered structure is composed of Bi$_2$S$_4$ layers, Bi$_2$O$_2$ layers and SO$_4$. The fluorite-type Bi$_2$O$_2$ layer is one of the well-known layers in the *I*4/*mmm* or *P*4/*nmm* tetragonal layered structures. In fact, Bi$_2$O$_2$SO$_4$ is a known compound. The rock-salt-type Bi$_2$S$_4$ layer structure has been found only in ReOBiS$_2$ (Re = Ce, Gd and Dy) with the *P*4/*nmm* space group.[19] Figure 2d is a schematic image of the Bi$_2$S$_4$ layer projected along the *c* axis. The Bi and S atoms form square lattices. When all the sites are fully occupied, the atomic composition should be Bi$_6$O$_8$S$_5$, namely Bi$_6$O$_4$S$_4$(SO$_4$). If the occupancy of the SO$_4$ site is 50 %, the composition is Bi$_4$O$_4$S$_3$. Therefore, we speculate that there are defects at the SO$_4$ site in our Bi$_4$O$_4$S$_3$ superconducting sample, because defects of molecules or ions at the interlayer site are often observed in layered compounds.

Finally, we discuss the band structure of the present material. It is found that Bi$_4$O$_4$S$_3$ ($x$ = 0.5) is metallic while Bi$_6$O$_8$S$_5$ ($x$ = 0) is band insulator with Bi$^{3+}$. Figure 3 shows the band structure for $x$ = 0.5 (Bi$_4$O$_4$S$_3$) obtained using the Wien2k package adopting the lattice structure obtained as above.[20] The Fermi level lies within the bands which mainly originate from the Bi 6p orbitals. In particular, the Fermi level is just on the peak position of the partial density of states of the Bi 6p orbital within the BiS$_2$ layer. Focusing on the contribution from the BiS$_2$ layer, it is found that the band structure near the Fermi level consists of p$_x$ and p$_y$ orbital contributions, which in itself have quasi-one-dimensional nature, but mixes around the R point to result in a two dimensional band structure. The width of these Bi p bands is about 3eV, which is comparable to the width of the 3d bands of the high-$T_c$ cuprates or the iron-based superconductors, and this, along with the low dimensional nature of the band structure suggests that electron correlation effects may be playing some role in the occurrence of superconductivity.

As reported here, superconductivity in the new layered superconductor Bi$_4$O$_4$S$_3$ is realized within the BiS$_2$ layers upon electron doping. By changing the spacer layers, we will be able to obtain various BiS$_2$-based new superconductors. We believe that the BiS$_2$-based family will open a new field in physics and chemistry for low-dimensional superconductors as have cuprate and Fe-based superconductors.

**Methods**

Polycrystalline samples of $Bi_4O_4S_3$ were prepared using a conventional solid state reaction method. All the chemicals were purchased from Kojundo Chemical Lab. At first, $Bi_2S_3$ was prepared by reacting Bi (99.99% grain) and S (99.9% grain) in an evacuated quartz tube above 500 ºC. Starting materials of $Bi_2S_3$, $Bi_2O_3$ (99.9% powder) and S were ground, pelletized, sealed into an evacuated quartz tube and heated at 510 ºC for 10 h. The product was well-ground, pelletized and annealed at 510 ºC for 10 h in an evacuated quartz tube again. In the case that the $Bi_4O_4S_3$ sample was heated above 550 ºC, $SO_2$ gas was produced. If you try to synthesize this material, please be careful not to heat the sample sealed in an evacuated tube at high temperature to avoid explosion.

Temperature dependence of magnetic susceptibility from 10 to 2 K after both zero-field cooling (ZFC) and fieled cooling (FC) was measured using a superconducting quantum interference device (SQUID) magnetometer with an applied magnetic field of 9.3 Oe. Temperature dependence of resistivity from 300 to 2 K was measured using the four terminals method. The $T_c^{onset}$ was defined to be a temperature where the resistivity deviates from the linear-extrapolation of normal-state resistivity as shown in the inset of fig. 1b.

X-ray diffraction (XRD) pattern was collected by RIGAKU x-ray diffractometer with Cu-Kα radiation using the 2θ-θ method. The diffraction peaks were indexed by the software X-Cell[21] of Accelrys, Inc. The initial model of the crystal structure was derived from the software TOPAS of Bruker AXS, Inc. The structure was refined by the Rietveld-analysis program Materials Studio Reflex of Accelrys, Inc. Geometry optimizations were carried out using the density functional theory (DFT) methods with the program Materials Studio CASTEP[22] of Accelrys, Inc. The GGA (generalized gradient approximation)–PBEsol (Perdew-Burke-Ernzerhof for solids) exchange-correlation functionals[23] and ultrasoft pseudopotentials[24] were employed. The energy cutoff for the plane-wave basis set was 380.0 eV. The Monkhorst-Pack grid separation[25] was set to approximately 0.04 Å$^{-1}$. The lattice parameters were set to the experimental values, and the atomic positions were optimized to minimize the total energy. The schematic images of the crystal structure were depicted using VESTA.[26]

Band structure of $Bi_4O_4S_3$ shown in Fig.3 was calculated using the WIEN2k package.[20] GGA-PBEsol exchange-correlation functional was adopted.[23] 512 k-points and $RK_{max}$ = 7.0 were taken. The spin-orbit interaction was omitted.

Acknowledgements

The authors would like to thank Dr. S. J. Denholme, Dr. T. Yamaguchi (National Institute for Materials Science) and Dr. H. Takatsu (Tokyo Metropolitan University) for their experimental helps and fruitful discussion. This work was partly supported by Grant-in-Aid for Scientific Research (KAKENHI) and JST-EU-JAPAN project on superconductivity.

Author Contributions

Y.M., H.I. and O.M. synthesized $Bi_4O_4S_3$ superconducting samples, performed XRD, and measured magnetic properties. H.F. and Y.G. performed crystal structure analysis. Y.M., S.D. and Y.T. measured transport properties. K.S., H.U. and K.K. performed band calculations. Y.M. prepared the manuscript with input from all authors.

Author Information

Correspondence and requests for materials should be addressed to Y.M. (mizugu@tmu.ac.jp).

Table 1. **Structural parameters of $Bi_6O_4S_4(SO_4)_{1-x}$ ($x = 0$) obtained by Rietveld refinement.**

| site | $x$ | $y$ | $z$ | occupancy |
|---|---|---|---|---|
| Bi1 | 0 | 0 | 0.0576(1) | 1 |
| Bi2 | 0 | 0 | 0.2082(1) | 1 |
| Bi3 | 0 | 0 | 0.3810(1) | 1 |
| S1 | 0 | 0 | 0.1410(3) | 1 |
| S2 | 0 | 0 | 0.2868(4) | 1 |
| S3 | 0 | 0 | 0.5 | 1 |
| O1 | 0 | 0.5 | 0.0884(fixed) | 1 |
| O2 | 0 | 0.3053(fixed) | 0.4793(fixed) | 0.5 |

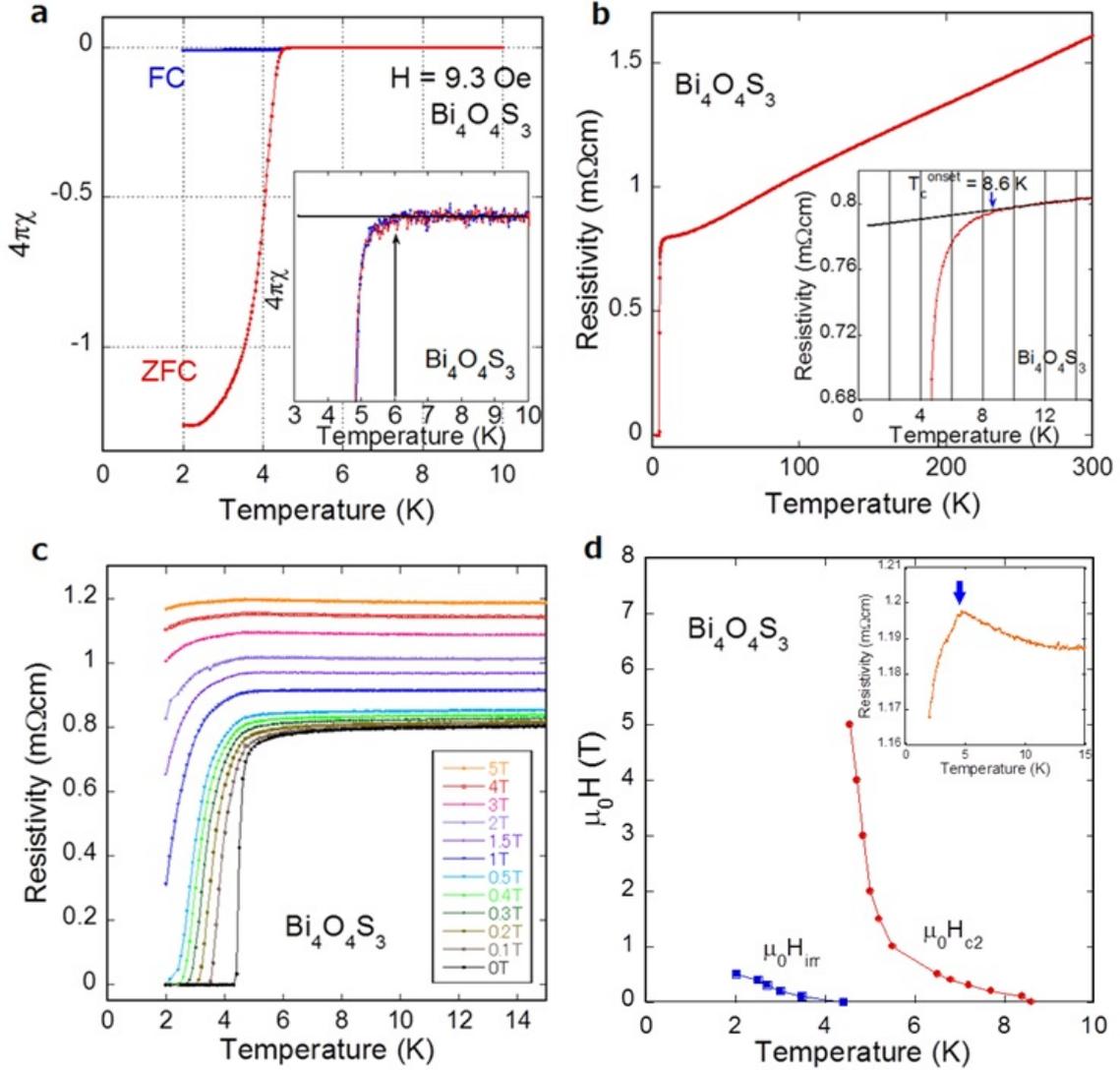

Fig. 1. **Superconducting properties of $Bi_4O_4S_3$.** **a**. Temperature dependence of magnetic susceptibility for $Bi_4O_4S_3$. The inset shows an enlargement of onset of the superconducting (diamagnetic) signal. **b**. Temperature dependence of resistivity for $Bi_4O_4S_3$. The inset is an enlargement of onset of the superconducting transition. **c**. Temperature dependence of resistivity for $Bi_4O_4S_3$ under magnetic fields up to 5 T. **d**. Magnetic field-temperature phase diagram in which the onset and zero-resistivity temperatures are plotted. The inset displays an enlarged superconducting transition of the temperature dependence of resistivity at 5 T.

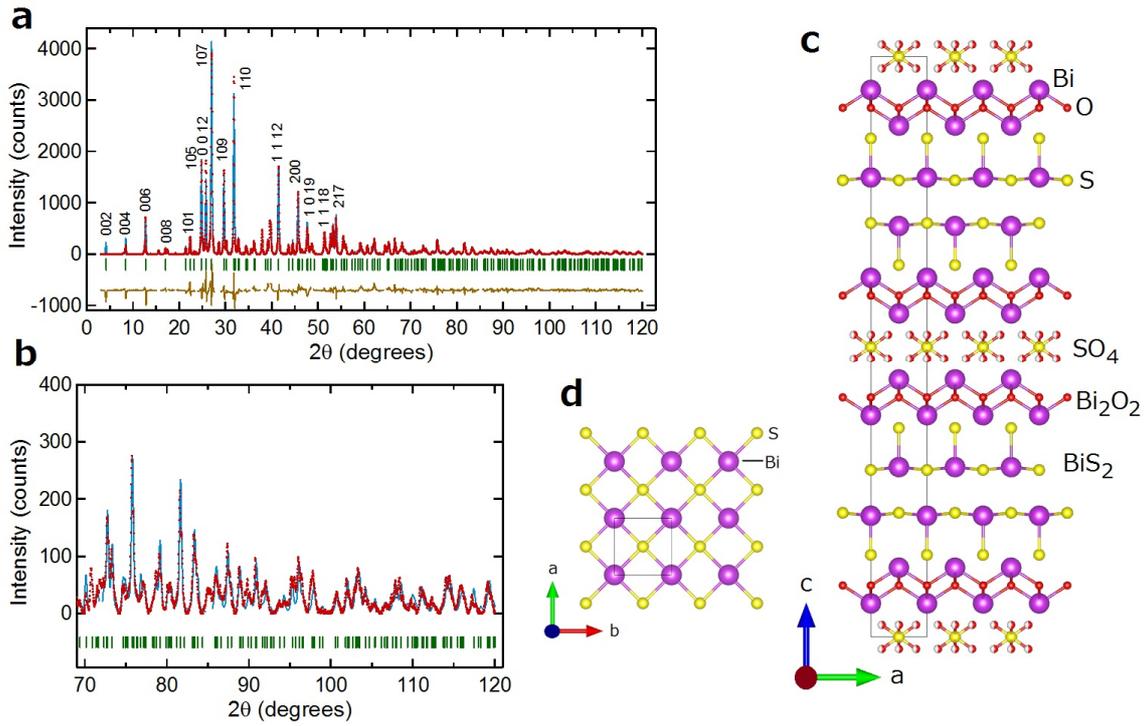

Fig. 2. **Crystal structure analysis of new superconductor Bi$_4$O$_4$S$_3$.** **a**. XRD pattern with the result of the Rietveld refinement. The displayed numbers are the Miller indices. **b**. XRD pattern at high angles. **c**. Schematic image of the obtained crystal structure. Purple, yellow and red circles indicate Bi, S and O atoms, respectively. The occupancy of the O2 site (near $z = 0$ and 0.5) is 0.5. The chemical composition can be described as Bi$_6$O$_4$S$_4$(SO$_4$)$_{1-x}$ when defects of SO$_4$ ions exist. Bi$_4$O$_4$S$_3$ corresponds to $x = 0.5$ where the site occupancies of S3 and O2 in Table 1 are 0.5 and 0.25, respectively. **d**. Schematic image of the BiS$_2$ square lattice (*ab* plane).

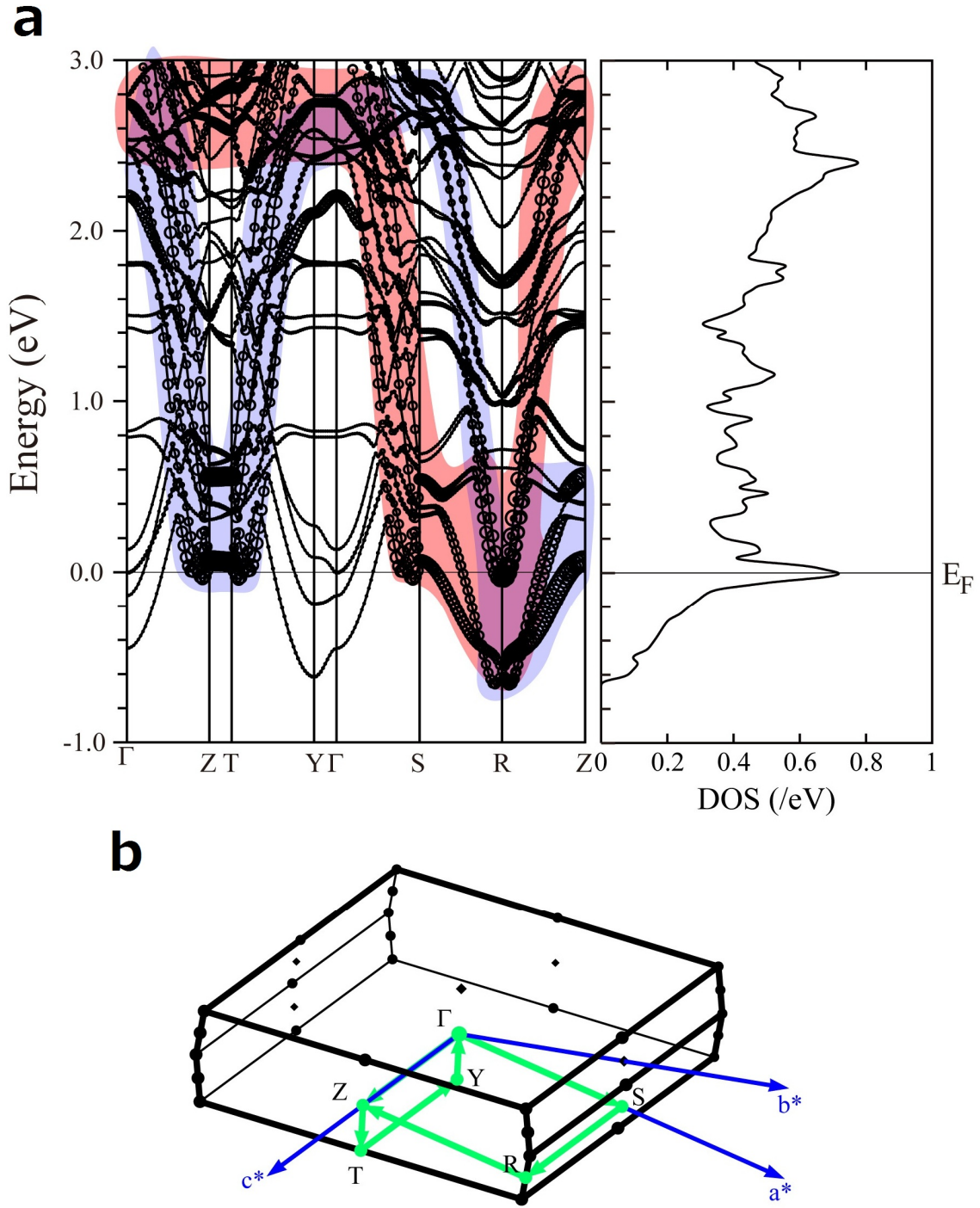

Fig. 3. **Band structure and Brillouin zone of $Bi_4O_4S_3$.** a. Left panel: the band structure for $x = 0.5$. The radius of the circles represent the weight of the Bi 6p orbitals within the $BiS_2$ layer. The blue and red hatches indicate the bands having mainly $p_x$ and $p_y$ characters, respectively. Right panel: the partial density of states of the Bi 6p orbital within the $BiS_2$ layer. b. The Brillouin zone is shown.

13